\documentstyle[prl,twocolumn,aps,psfig]{revtex}

\draft
\begin{document}
\twocolumn[\hsize\textwidth\columnwidth\hsize\csname@twocolumnfalse%
\endcsname

\title{Quantum Antiferromagnets in a Magnetic Field}

\author{Daniel Loss and B. Normand}

\address{Institute of Physics, University of Basel, Klingelbergstrasse 82, 
CH-4056 Basel, Switzerland. }

\date{\today}

\maketitle

\begin{abstract}

	Motivated by recent experiments on low-dimensional quantum magnets 
in applied magnetic fields, we present a theoretical analysis of their 
properties based on the nonlinear $\sigma$ model. The spin stiffness and 
a 1/N expansion are used to map the regimes of spin-gap behavior, 
predominantly linear magnetization, and spin saturation. A two-parameter 
renormalization-group study gives the characteristic properties over the 
entire parameter range. The model is relevant to many systems exhibiting 
Haldane physics, and is applied here to the two-chain spin ladder compound 
CuHpCl. 

\end{abstract}

]


	The importance of low-dimensional spin systems in revealing 
fundamentally new quantum mechanical properties has been recognized since 
Haldane's conjecture \cite{rh} concerning the effects of quantum fluctuations 
in integer- and half-integer-spin antiferromagnetic (AF) chains. We begin 
with the experimental observation that the magnetization curves of some such 
materials, thought to be prototypical of the extreme quantum limit, are in 
fact remarkably classical. As examples, we refer here to the Haldane 
($S$ = 1) chain NENP \cite{ragksi}, and primarily to the two-chain $S = 
{\textstyle \frac{1}{2}}$ ladder CuHpCl \cite{rcclpmm,rhrbt}, which show 
regimes of linear magnetization whose gradient is the (classical) N\'eel 
susceptibility. 

	With the goal of understanding such behavior, we consider the 
quantum AF system in an external magnetic field using the nonlinear 
$\sigma$ model (NLsM). This widely-applied treatment is in fact 
semi-classical, being truly valid only in the limit of large spin $S$, 
but has in the past formed the basis for many fundamental deductions 
concerning the quantum limit. We will demonstrate here its applicability 
for effectively integer-spin 
quantum systems in appreciable magnetic fields, and provide justification 
of this result in terms of the suppression of quantum fluctuations by the 
field. For the purposes of developing the present theoretical analysis, 
we will concentrate on the best-characterized sample in recent literature, 
Cu$_2$(1,4-diazacycloheptane)$_2$Cl$_4$ (CuHpCl) \cite{rcclpmm,rhrbt},

	The Hamiltonian for the ladder system in a magnetic field ${\bf b}
= \tilde{g} \mu_B{\bf B}$ may be written as 
\begin{equation}
H = \sum_{i, m=1,2} [J {\bf S}_{m,i} {\bf \cdot S}_{m,i+1} + J^{\prime} 
{\bf S}_{1,i} {\bf \cdot S}_{2,i} + {\bf b \cdot S}_{m,i}],  
\label{esh}
\end{equation}
where $J$ is the intrachain exchange interaction and $J^{\prime}$ the 
interchain, or ladder ``rung'', interaction. The derivation of the 
NLsM in the presence of a magnetic field is presented in Ref. 
\onlinecite{ral}. For $N_x$-site chains with periodic boundary conditions 
along $x$, in the geometry shown in Fig. 1, the resulting model has
$J_x = J$, $J_y = {\textstyle \frac{1}{2}} J^{\prime}$, ${\bf b} = (0,0,b)$, 
and ${\bf S}_{m,N_x+i} = {\bf S}_{m,i}$. There are two key 
points. First, the spin is written in terms of slowly-varying, 
orthogonal, staggered and uniform components as ${\bf S}_{m,i} \simeq 
S[(-1)^{i+m} {\bf n}_{m,i} + a {\bf l}_{m,i}]$, where $a$ is the lattice 
constant, and the fluctuations ${\bf l}$ about the staggered configuration 
must be integrated out subject to the orthogonality constraint ${\bf n \cdot 
l} = 0$. Second, the full Euclidean action in space and inverse temperature, 
${\cal S}_E = {\cal S}_{B} + \int_{0}^{\beta} d \tau H$, contains in addition 
a Berry-phase term 
${\cal S}_{B} = i \frac{S}{a^2} \int d \tau dx 
[ - a {\bf l \cdot} ({\bf n} \wedge {\dot{\bf n}})] + 4 \pi i S 
(P_1 + P_2)$. 
$P_1 = P_2 = {\textstyle \frac{1}{4\pi}} \int d \tau dx ({\bf n} \wedge 
{\dot{\bf n}}) {\bf \cdot} \partial_x {\bf n}$ separate to give simply 
the Pontryagin index on each chain when $\partial_y {\bf n} = 0$, as is 
the case in a ladder of only two chains. The last term in ${\cal S}_{B}$ 
is thus $i (4 \pi P_1) 2 S$, demonstrating that the system will have 
integer-spin characteristics for any value of $S$, and the topological 
term may be ignored \cite{rk,rs}. 

	The action for the quasi-one-dimensional ladder system, in 1+1 
Euclidean dimensions denoted by $\mu$, is then 
\begin{equation}
{\cal S}_E = \frac{1}{2 g} \int_{\tau,x} [(\partial_{\mu} {\bf n})^2
 - ({\bf b}^2 - ({\bf n \cdot b})^2) + 2i {\bf b \cdot} {\bf n} \wedge 
\dot{\bf n}], 
 \label{ese}
\end{equation}
where $g = (2 / N_y S) \sqrt{\bar{J} / J_x}$ is the bare coupling constant, 
$\bar{J} = J_x + {\textstyle \frac{1}{2}} J_y$, and the integral over 
$\tau$ is to upper limit $L_T = c \beta$, with $c = 2 S a \sqrt{J_x 
\bar{J}}/\hbar$ ($\hbar \rightarrow 1$) the effective spin-wave velocity. 
We have left explicit the number $N_y$ of chains in the ladder; for the 
compound under consideration $J_y > J_x$ and $N_y = 2$, giving an effectively 
rigid rung coupling. 
The form of the NLsM in an external field given by the second term 
in Eq. (\ref{ese}) has been derived previously \cite{ral}, and its 
implications considered in the low-field limit. In what follows we will 
examine its effects for arbitrary fields.


	To gain initial insight into the effect of the magnetic field, we 
consider the spin stiffness of the ladder system \cite{rlm}. Taking the 
staggered spin configuration to be subject 
to a twist $\theta$ in the plane normal to the applied field (Fig. 1), we 
calculate the free energy $F(b,\theta)$ to one-loop order in the important 
out-of-plane spin fluctuations, and deduce the spin stiffness from $\rho_s 
= \frac{1}{2} c L \left. \frac{\partial^2 F}{\partial \theta^2} 
\right|_{\theta = 0}$. 
\begin{equation}
\rho_s = \rho_s^0 \left[ 1 - \frac{g}{L L_T} \sum_{\bf k} \frac{1}{{\bf 
k}^2 + (b/c)^2} \right],
\label{ess}
\end{equation}
where $\rho_s^0 = c/2g$ is the classical value and the sum provides both 
quantum and thermal (through the finite ``length'') corrections to first 
order. 

\begin{figure}[hp]
\centerline{\psfig{figure=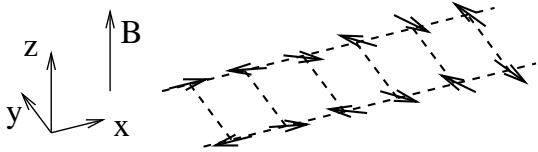,height=2.0cm,angle=270}}
\medskip
\caption{Representation of two-chain ladder system, showing 
directions of staggered moment, twist and applied field.}
\end{figure}

	We restrict the analysis to the low-temperature, or 
``quantum''  case $L_T > L$. Evaluating the summation between spatial 
limits $\pi/L$ and $\pi/a$, and introducing the ``magnetic length'' $L_m$ 
as $\pi/L_m = b/c$, 
\begin{equation}
\rho_s = \rho_s^0 \left[ 1 + \frac{g}{4 \pi} \ln \left( \frac{(a/L)^2 + 
(a/L_m)^2}{( 1 + (a/L_m)^2)} \right) \right].
 \label{efsse}
\end{equation}
$L$ in Eq.(\ref{efsse}) may be regarded as the correlation length $\xi$, 
beyond which segments of the ladder behave independently, and can be 
computed from $\rho_s = 0$. In the weak-field limit ($L_m \rightarrow 
\infty$) one recovers the result \cite{rlm} $\xi_0 = a e^{2 \pi / g} = a 
e^{\alpha \pi S}$, where $\alpha = \sqrt{J_x / \bar{J}}$. 
The general solution is $\xi(B) = \xi_0 / \sqrt{1 - ( L_{m}^* / L_m )^2}$, 
where $L_{m}^* = a \sqrt{e^{4 \pi / g} - 1}$ gives the critical field 
$B^*$ at which the correlation length diverges. For fields $B < B^*$, 
the finite correlation length may be written as $\xi(B) = a e^{\alpha \pi 
\tilde{S}}$, where $\tilde{S} = S [ 1 - g / 4 \pi \ln ( 1 - ( L_{m}^* / L_m 
)^2 )]$ is a growing value of the effective spin. For $B > B^*$, the field 
enforces a quasi-long-ranged correlation throughout the system \cite{ra}, 
and it is most convenient to write the spin stiffness as $\rho_s = 
\rho_{s}^0 [ 1 - g / 4 \pi \ln ( 1 + ( L_m / a )^2 )]$, which 
recovers the bare value as $B \rightarrow \infty$. Finally, the divergence 
of the correlation length at $B^*$ corresponds to the closing of the gap 
$\Delta$ to spin excitations according to $\Delta \propto \sqrt{1 - 
( B / B^*)^2}$. This situation is summarized in Fig. 2. 

	However, if the $B$ derivative is taken 
of the free energy in order to compute the magnetization, the 
resulting terms do not yield the required zero result in the spin gap regime.
This indicates the breakdown of the approximation implicit in deriving $F(b,
\theta)$ that the field be large on the scale of in-plane fluctuation energies 
$|\dot{\phi}|$. Thus the spin stiffness analysis, while qualitatively 
revealing of the behavior of the system, is not reliable at low fields.


	The situation in the weak-field regime may be addressed by 
a $1/N$ expansion \cite{rp}, which is expected to be appropriate in 
describing spin-gap phases. The staggered spin ${\bf n}$ is taken to 
exist in an $N$-dimensional spin space, in which only the component $n_z$ 
is selected by the field. The relevant parts of the $O(N)$ action 
are 
\begin{equation}
{\cal S}_E = \frac{1}{2 g} \int d {\bf x} (\partial_{\mu} {\bf n})^2
 - \bar{b}^2 ( 1 - n_{z}^2 ) - i \lambda ( {\bf n}^2 - 1 ), 
 \label{es1n}
\end{equation}
where $\bar{b}$ denotes $b/c$ and the constraint that ${\bf n}$ 
have unit magnitude is made explicit with the Lagrange multiplier $i 
\lambda$, whose saddle-point value is given by 
\begin{equation}
\frac{1}{g} = (N-1) \sum_{\bf k} \frac{1}{{\bf k}^2 + i \lambda} + 
\sum_{\bf k} \frac{1}{{\bf k}^2 + i \lambda + \bar{b}^2} ,
\label{elc}
\end{equation}
in which the $B$-field term is found to appear only at $O(1/N)$. $i \lambda$ 
functions as a mass, or cutoff term in momentum integrations, and is thus an 
upper lengthscale for cooperative processes in the system, or simply a 
correlation length (inverse excitation gap). Writing the saddle-point 
solution as $i \lambda = c^2 \pi^2 / \xi(B)^2$ and carrying out the summation 
at low $T$ gives an expression analogous to 
$\rho_s = 0$ emerging from Eq.(\ref{efsse}).
 
\begin{figure}[hp]
\centerline{\psfig{figure=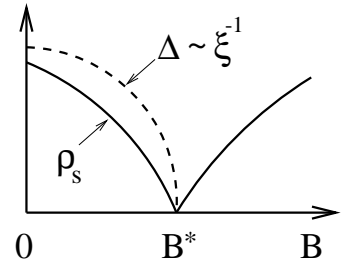,height=3.5cm,angle=270}}
\medskip
\caption{Schematic depiction of behavior of correlation length or spin 
gap, and of spin stiffness, with applied field. }
\end{figure}

At weak fields ($\xi \ll L_m$) one finds $\xi \simeq a e^{2 \pi / N g} ( 1 + 
(a^2/NL_{m}^2) e^{4 \pi / g}$, a $O(1/N)$ correction to a result differing 
from the previous one by a power of $1/N$ in the exponent. The 
result $\frac{\partial}{\partial b} \left( \frac{c^2 \pi^2}{\xi^2} \right) 
= - \frac{2 b}{N}$ ensures both that the corresponding magnetization 
contribution from the ${\bf k}$ summation terms in $F$ is identically zero 
to $O(1/N)$, and that the constant term $b^2 + c^2 \pi^2 / \xi^2$ yields 
$2b (1 - 1/N)$. It is clear that the behavior required of a gapped 
system is returned in the weak-field regime only on making the 
well-recognized identification (motivated by comparison with 
renormalization-group results \cite{rp}) $N \rightarrow N - 2$, and by
returning to the physical situation $N = 3$. Then the magnetization is 
indeed zero, and the saddle-point solution for $\xi (B)$ becomes precisely 
that deduced from the spin stiffness analysis, with the same critical field 
$B^*$. With this replacement caveat we thus obtain from these two approaches 
a consistent picture of both weak- and strong-field regimes. 


	To quantify the regimes of validity of the foregoing analyses, we 
consider next a renormalization-group (RG) approach to the NLsM in an applied 
field. Taking the model in the form 
\begin{equation}
{\cal S}_E = \frac{1}{2 g} \int d {\bf x} [ (\partial_{\mu} {\bf n})^2
 - \bar{b}^2 (1 - n_{z}^2) + 4 i \bar{b} n_x \dot{n}_y ]  
 \label{esrg}
\end{equation}
and transforming to variables $\phi$ and $\sqrt{g} \sigma_z = n_z$ 
representing respectively the in- and out-of-plane fluctuations, the latter 
in a form suitable for perturbative expansion in the coupling constant $g$, 
we obtain
\begin{equation}
{\cal L}_E = \frac{1}{2 g} \left( A - \bar{b}^2 \right) - \frac{1}{2} 
\sigma_z ( - \partial_{\mu}^2 + \bar{b}^2 - A) \sigma_z + O(g). 
\label{elrg}
\end{equation}
$A$ denotes the in-plane terms $(\partial_{\mu} \phi)^2 + 2 i \bar{b} 
\dot{\phi}$, which because $\phi(\tau,x)$ is assumed to vary slowly can be 
taken to be a small constant (no fast Fourier modes) in the momentum shell 
$\gamma \pi / a < |{\bf k}| < \pi / a$ ($\gamma \rightarrow 1$). 
Performing the integral and expanding in $A$, the form of Eq.(\ref{elrg}) 
is recovered with coefficients $g(a')$ and $\bar{b}(a^{\prime})^2$ given 
by partial traces. Evaluation of these and differentiation with 
respect to the flow parameter $l = \ln (a^{\prime}/a)$, leads to the coupled 
RG equations
\begin{equation}
\frac{d g}{d l} = \frac{g^2}{2 \pi} \frac{1}{ 1 + \bar{\beta}^2}, \;\;\;\;
\frac{d \bar{\beta}^2}{d l} = 2 \bar{\beta}^2 - 
\frac{g^2}{2 \pi} \ln \left( 1 + \bar{\beta}^2 \right), 
\label{erge} 
\end{equation}
in which $\bar{\beta} = a^{\prime} \bar{b} (a^{\prime})$. These new RG 
equations possess a variety of interesting limiting cases, whose detailed 
study we defer to a future publication \cite{rnkl}. 

	For the present purposes, we concentrate on the fixed points to 
obtain a qualitative picture of the RG flow diagram, and on the consequences 
for the magnetization. (i) Seeking a fixed point by weak-field expansion 
around $\bar{\beta}_* = 0$, we find $dg / dl \simeq g^2 / 2 \pi$ and thus 
$d \ln \bar{\beta}^2 / dl = 2 - d \ln g / dl$, 
which may be solved to yield
\begin{equation}
\frac{g_0}{g} = 1 - \frac{g_0}{2 \pi} l, \;\;\;\;\;\; \bar{\beta} = 
\bar{\beta}_0 e^l \left( 1 - \frac{g_0}{2 \pi} l \right)^{1/2}. 
\label{ergfp} 
\end{equation}
The fixed point $(g_{*},\bar{\beta}_{*}) = (\infty,0)$ is clearly stable if 
the flow is stopped at $l_* = 2 \pi / g_0$. The system will flow to this 
strong-coupling regime if the starting value $\bar{\beta}_0$ is sufficiently 
small. The lengthscale ${\cal L}_* = a e^{l_*}$ at which the flow stops may 
be compared with the spatial and thermal dimensions ${L,L_T}$ of the system 
to calculate directly the effects of finite size and temperature \cite{rnkl}. 
(ii) At strong fields $(\bar{\beta} 
\rightarrow \infty)$, $dg/dl = 0$, or $g = g_0$, indicating that the coupling 
is not renormalized, and $d \ln \bar{\beta}^2 = 2 dl$, from which 
it follows that $\bar{b}(l) = \bar{b}_0$, {\it i.e.} neither is the field. 

	Numerical solution of Eqs.(\ref{erge}) leads to the 
flow diagram in Fig. 3. The regime (ii) of strong initial field 
may be termed the weak-coupling situation, where $g$ and $\bar{b}$ are 
weakly renormalized to finite values. Here the perturbation theoretic 
approach is consistent and the weak coupling corresponds to deconfinement 
of the excitations on the lengthscale $L_m$ set by the field. The regime 
(i) is the strong-coupling condition, of confinement of (gapped) 
excitations, where the assumption of small $g$ in the derivation proves 
to be inconsistent. However, in this regime one may deduce the critical 
lengthscale ${\cal L}_*$ governing the behavior of the system and, as we 
will show below, that the magnetization is zero (inset Fig. 3). The critical 
starting field separating the two regions is $\bar{b}_* = 0.46$. The 
properties of the two regimes may be illustrated by considering the 
correlation length in each. $\xi$ is a physical quantity and does not 
change under the RG flow, meaning that $d \xi / d a^{\prime} = 0$, whence  
\begin{equation}
\frac{\partial g}{\partial l} \frac{\partial \xi}{\partial g} + 
\frac{\partial \bar{\beta}^2}{\partial l} \frac{\partial \xi}{\partial 
\bar{\beta}^2} + \xi = 0.
\label{ecli}
\end{equation}
(i) At small $\bar{\beta}^2$, $- \partial \xi / \xi \simeq 2 \pi \partial 
g / g$, which has solution $\xi_0 = \xi e^{2 \pi (1/g_0 - 1/g)}$. Under the 
reasonable assumption that $\xi \rightarrow a$, the bare lattice constant, 
as the system flows to the strong-coupling limit ($1/g_0 \rightarrow 0$), we 
recover the expected result $\xi_0 = a e^{2 \pi / g_0}$ (see above) for the 
finite physical correlation length. (ii) For large $\bar{\beta}^2$, $- 2 
\partial \xi / \xi \simeq \partial \bar{\beta}^2 / \bar{\beta}^2$ leads to 
$\xi_0 = \xi {\cal L}/a$ ($\bar{b}_0$ invariant) under the RG flow, and we 
see that for any assumed finite $\xi$, the bare correlation length $\xi_0$ 
is the system size $L$ or $L_T$. 

\begin{figure}[hp]
\centerline{\psfig{figure=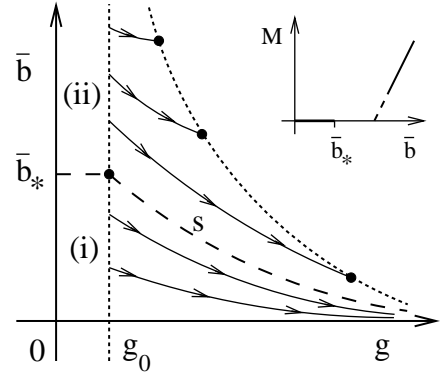,height=5cm,angle=270}}
\medskip
\caption{RG flow diagram for $g$ and $\bar{b}$. Strong- and weak-coupling 
regimes are separated by separatrix s.}
\end{figure}


	We return to the experimental motivation for the above analysis, 
and compute the magnetization $M = \partial F / \partial B$ of the model 
over the full field range,  
\begin{equation}
M = - \tilde{g} \mu_B N_x N_y (b / 4 \bar{J}) + M_d + M_f .
\label{em}
\end{equation}
M contains contributions linear in $B$ from the quadratic term (\ref{es1n}), 
$M_d$ from the dynamical term in the trace ({\it cf.} (\ref{ess})) due to 
fluctuations of ${\bf n}$ out of the plane perpendicular to ${\bf B}$, and 
$M_f$ from in-plane fluctuation terms 
$\dot{\phi}$. The last give a sawtooth form leading to magnetization steps
\cite{rnkl}, a finite-size effect which will not be considered further 
here. While the linear term is always present, we have shown that below a 
threshold field $B^*$, where the system has a spin gap, it is cancelled by 
the corresponding correlation-length term. Above an upper threshold $B_{c2}$, 
the magnetization will saturate at the value $M_s = \tilde{g} \mu_B S N_s$ 
($N_s = N_x N_y$), and this effect is not contained in the (large-$S$) model 
as applied. Systematic inclusion of an additional total spin constraint is 
possible \cite{rnkl}, but to compress the analysis we will here apply $M_s$ 
as a simple cutoff. The ``dynamical'' contribution is given consistently 
to lowest order in the small parameter $c/b$, and in limit of low $T$, by the 
constant $M_d = {\textstyle \frac{1}{2}} N_x \tilde{g} \mu_{B}$. Next-order 
corrections involving the spin excitations have the form $M_{d}^{\prime} 
\sim B \ln (B / \bar{J})$. 

	Specializing to the two-chain ladder material CuHpCl, the exchange 
constants deduced from the magnetization and susceptibility 
\cite{rcclpmm,rhrbt} are $J^{\prime}$ = 13.2K and $J$ = 2.4K, 
whence $\bar{J}$ = 13.3T and $J_x$ = 3.6T. Taking the 
simplest case of constant $M_d$, and the lower critical field $B_{c1}$ for 
onset of magnetization where $M(B_{c1}) = 0$, we obtain $B_{c1} = 
\bar{J}/\tilde{g}$ = 6.6T. The saturation field $B_{c2}$ is given from  
$M(B_{c2}) = M_s$ as $B_{c2} = 4 S / \tilde{g} \bar{J}$ = 13.3T. These 
values are in remarkably good agreement with a linear extrapolation of the 
magnetization data at lowest temperature in Ref. \onlinecite{rcclpmm}, 
where $B_{c1}$ = 
6.8T and $B_{c2}$ = 13.7T. Such an extrapolation appears closer to the data 
than the predictions \cite{rhpl} of a repulsive boson model. We note further 
that the gradient of the linear magnetization is precisely the classical 
N\'eel susceptibility $\chi_{AF}^{\perp} = (\tilde{g} \mu_B)^2 / (4 
\bar{J} a^2)$ per unit volume, a result in itself remarkable for an 
AF in the quantum limit. The computed magnetization is shown in Fig. 4, 
where the dashed line indicates the validity limit of the calculation. 

\begin{figure}[hp]
\centerline{\psfig{figure=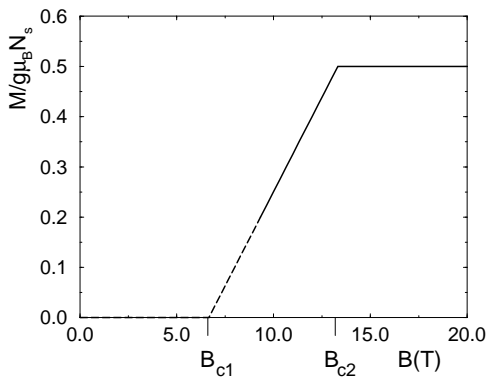,height=5.5cm,angle=270}}
\medskip
\caption{Computed magnetization of ladder system CuHpCl.}
\end{figure}

	The predicted magnetization shows a surprising degree of accuracy 
for a model strictly valid only in the limit of large $S$. We note that the 
NLsM has recently been used with considerable success also for small systems 
with small $S$ \cite{rcl}. Its accuracy in the current context may be 
ascribed in part to the suppression of quantum fluctuation effects by the 
field, as shown above, such that in the high-field 
regime, which covers much of that beyond $B_{c1}$, the behavior of a system 
with no topological term is rather classical. Weak interladder interactions 
causing the real material to display 3d order at intermediate fields will 
also enhance classical behavior. 

	Further perspective on the above results is given by computing 
the critical field $B^*$ from the spin stiffness and $1/N$ treatments. We 
deduce from $L_{m}^*$ that $B^* = 0.32\bar{J} / \tilde{g}$ = 2.1T, a value 
rather lower than $B_{c1}$ above, and emphasize that in the regime between 
these values, as also seen in the RG analysis, the treatment is not reliable. 
In the RG approach, the magnetization 
\begin{equation}
M = \frac{\tilde{g} \mu_B a}{c} \frac{\partial F}{\partial \bar{\beta}_0} 
  = \frac{\tilde{g} \mu_B a}{c} \left( \frac{\partial \bar{\beta}}{\partial 
\bar{\beta}_0} \frac{\partial F}{\partial \bar{\beta}} + \frac{\partial 
g}{\partial \bar{\beta}_0} \frac{\partial F}{\partial g} \right), 
\label{emrg}
\end{equation}
from which we see in the small-$B$ regime, where $\partial g / \partial 
\bar{\beta}_0 = 0$ and $\partial \bar{\beta} / \partial \bar{\beta}_0 
= e^l \sqrt{1 - g_0 l / 2 \pi}$, that for any $F(\bar{b})$ analytic in 
$\bar{b}$, $M(B_0) \sim \sqrt{1 - g_0 / 2 \pi \ln ({\cal L} / a)} 
\rightarrow 0$ as ${\cal L} \rightarrow {\cal L}^*$. Thus the magnetization 
vanishes in the strong-coupling limit as required. From the numerical 
solution, the field scale $\bar{b}_*$ gives $B_*$ = 1.6T, in reasonable 
agreement with the value $B^*$ above. We have already shown that the regime 
approaching the physical $B_{c1}$ lacks a suitable calculation scheme. 
Physically, the problem is one of determining the effects of quantum 
fluctuations of the spin system when the spin gap is weakened by the applied 
field, and the quantum disordered phase thus made less robust. We have 
detailed above the lowest-order predictions of the NLsM for the closing 
of the gap, and expect these to be appropriate when the gap becomes 
smaller than other energy scales (quantum, thermal and finite-size 
fluctuations) in the system. 

	The foregoing analysis is not restricted to CuHpCl, but applies 
also to the $S = 1$ AF (``Haldane'') chain. NENP is considered to be a 
prototypical case for observation of the Haldane gap but for the 
complication of a large single-ion 
anisotropy. The present study predicts again the qualitative features of 
a gapped regime of zero magnetization, followed by approximately linear 
behaviour towards saturation (not achieved), as in experiment 
\cite{ragksi}. We will present elsewhere the quantitative 
aspect of this problem. There has been considerable recent interest in the 
possibility of magnetization plateaus in certain systems, and we observe 
that in the current model these may be expected, for example in $S > 1$ 
chains, when the field strength is such that the projected in-plane spin 
$S {\bf n}_{\perp}$ is of integer amplitude, leading to a gapped phase. 


	In summary, the nonlinear $\sigma$ model treatment reproduces well the 
behavior of quantum antiferromagnets in an external field. For effectively 
integer-spin systems, meaning those with a trivial topological term, this 
statement is valid even in the extreme quantum limit of low spin and low 
dimensionality, because the magnetic field acts to suppress quantum 
fluctuation effects. 


	We are grateful to the Swiss National Fund for financial support. 
BN wishes to acknowledge the generosity of the Treubelfonds.

\end{document}